\documentclass[12pt]{iopart}

\usepackage{amsfonts}
\usepackage{graphicx}

\newcommand{\eqref}[1]{(\ref{#1})}

\newcommand{\afgap}{\Delta}

\begin{document}
\author{R Gersch$^1$, J Reiss$^1$ and C Honerkamp$^2$}
\address{$^1$ Max-Planck-Institut f\"ur Festk\"orperforschung, 
  Heisenbergstra\ss e 1, D-70569 Stuttgart, Germany}
\address{$^2$ Institut f\"ur
  Theoretische Physik, Universit\"at W\"urzburg, Am Hubland, 
  D-97074 W\"urzburg, Germany}
\ead{r@roland-gersch.de}
\title[fRG for 1st-order phase transitions]{Fermionic functional renormalization group for
  first-order phase transitions: a mean-field model}
\begin{abstract}
First-order phase transitions in many-fermion systems are not detected in 
the susceptibility analysis of common renormalization-group (RG) approaches. 
Here we introduce a counterterm technique within the functional 
renormalization-group (fRG) formalism which allows access to all 
stable and metastable configurations. 
It becomes possible to study symmetry-broken 
states which occur through first-order transitions
as well as hysteresis phenomena. For continuous  
transitions, the standard results are reproduced. As an example, we study 
discrete-symmetry breaking in a mean-field model for a commensurate 
charge-density wave. An additional benefit of the approach is that
away from the critical temperature 
for the breaking of discrete symmetries large 
interactions can be avoided at all RG scales.  
\end{abstract}
\maketitle

\section{Introduction}
\label{sec:intro}

Renormalization group (RG) methods are powerful systematic tools to 
investigate the emergence of the low-energy physics of many-particle
systems.  In the Wilsonian RG approach \cite{PhysRevB.4.3184,1994RvMP...66..129S,SalmhoferBook} 
for a many-fermion system, 
the quasiparticle excitations are integrated out successively with 
an energy scale decreasing toward the Fermi energy. 
This generates 
an energy-scale dependent effective interaction from which the 
leading interaction processes and correlations can be read off. Although
most implementations involve a number of approximations, the flow
of the vertex functions of the effective theory can be derived from an
exact RG equation for a generating functional \cite{2001PThPh.105....1S,1993PhLB..301...90W}. 
This powerful equation 
sparks the idea that a much more controlled approach to 
the low-energy phase than obtained previously is indeed possible.

In many applications for correlated electron systems one is primarily 
interested in changes of the ground state toward symmetry-breaking states. 
If there is such a symmetry change of the ground state, a certain bilinear 
$B_{ij}$ in the fermion operators $c_i^{(\dagger)}$, $c_j^{(\dagger)}$ with some quantum numbers $i$ and $j$ takes a nonzero anomalous (i.e. violating 
the orginal symmetry of the action) expectation value at a low RG scale. 
For a second order transition, this occurs continuously as RG scale or temperature are lowered. Then the tendency toward 
symmetry breaking is signaled in the RG flow 
by a divergence of a certain component 
of the effective interaction at a nonzero energy scale $\Lambda_c$. In 
correspondence with this, the zero-frequency susceptibility for the 
coupling to the bilinear operator $B_{ij}$ diverges in power-law form
at the same critical scale $\Lambda_c$. If we view the expectation value
$\langle B_{ij} \rangle $ as a bosonic field $\phi$ (e.g. by a 
Hubbard-Stratonovich transformation), the inverse susceptibility appears in the quadratic part of the action for the $\phi$-field. 
The symmetry breaking occurs at the
temperature or energy scale where this quadratic part changes its sign, i.e. where the effective potential for $\phi$ 
reverses its curvature at $\phi=0$ from positive to negative, 
such that the potential minimum occurs at a nonzero value of $\phi$. When the energy scale or temperature is lowered further, this minimum moves further away from zero and finally saturates unless collective fluctuations move it back to zero and restore the symmetric state (see, e.g., \cite{2004PhRvB..70l5111B}).

As indicated in this description, the passage from the symmetric state at 
high energy scales or temperatures to the symmetry-broken state at low 
energies is most easily described in a bosonic language, and
such methods are being developed 
(\cite{2006JPhA...39.8205S,2004PhRvB..70l5111B,krahlwetterich}). 
However, in many 
interesting examples in many-fermion systems it is not clear from the outset 
which 
type of boson or which fermion bilinear acquires a nonzero expectation 
value. Hence, a purely fermionic description 
is
a less biased approach, 
as it allows the system to develop its intrinsic correlations without 
any prejudice
of grouping certain fermions in pairs and treating these in a saddle-point
approach. Recently, some of us have presented a purely fermionic functional 
RG method based on \cite{2004PhRvB..70k5109K} which allows one to 
continuously flow into the symmetry-broken
state, requiring only an infinitesimal symmetry-breaking in the 
initial conditions. This approach has been shown to work correctly for 
models where mean-field solutions are exact, like the reduced 
Bardeen-Cooper-Schrieffer (BCS) model \cite{1957PhRv..108.1175B} 
or a similar charge-density-wave model with discrete-symmetry 
breaking \cite{2004PThPh.112..943S,2005EPJB...48..349G}. Extensions 
for more general models will follow.

Yet, these new fRG extensions face two problems. First, there is still an 
almost-divergence of the interactions at the critical scale. Large values 
of the effective interaction are problematic for the extension to 
more general models,  as they might cause a breakdown of the 
perturbative approach. A true divergence is avoided by the small 
explicit symmetry breaking put into the initial conditions. But in 
order to stay close to the symmetric model which 
one typically wants to study, some components of the interaction in the
symmetry-breaking channel still reach large values whose feedback on
other channels needs to be understood. While this problem may be solved
either by a compromise with an adequate size of the initial symmetry 
breaking or by understanding the potentially rich and interesting physics 
of the effect of these large interactions on the flow, a second problem is 
more fundamental.
The currently available fRG methods only test the stability of the symmetric 
phase {\em locally} around the origin in the space of anomalous expectation 
values. If there is no 
divergent susceptibility or no curvature change in the effective potential of 
a bosonic field, the fRG flow will remain finite and the anomalous expectation 
values will vanish continuously when the small initial symmetry-breaking is 
sent to zero. This in turn would be interpreted as stability of the symmetric 
phase and absence of spontaneous symmetry breaking. 
On the other hand, simple mean-field treatments of many models indicate 
first-order transitions into the symmetry-broken state. This occurs for 
example in the two-dimensional $t$-$t'$ Hubbard model near half filling
for the (commensurate) antiferromagnetic spin-density wave transition
\cite{reissrohemetzner,juliusdiss}
and in a model with a particular forward scattering for the transition into 
a state with broken fourfold symmetry of the Fermi surface
\cite{2005PhRvB..72c5114Y}. The fRG flow
for these parameters in the almost symmetric state finds an enhanced 
but not a diverging susceptibility, hence missing the symmetry change of 
the ground state. The fRG method simply does not see the 
thermodynamic-potential
minimum for nonzero order parameter if it is separated
from the symmetric minimum by a finite energy barrier,
as in the case of certain first-order transitions.
What is needed is a method that, 
intuitively described,
starts the flow further out in the symmetry-broken
range and which then flows into the correct minimum corresponding to a 
spontaneously broken state of a model {\em with a fully symmetric action}.

In this work, we present a counterterm extension of the fRG method which 
is able to find symmetry-broken states not captured by the 
treatment of \cite{2004PThPh.112..943S,2005EPJB...48..349G}. 
These would occur via first-order 
transitions upon varying temperature, 
chemical potential, 
or infrared cutoff.
The procedure invokes a flow differing slightly from Wilson's
momentum-shell RG.
The propagator of the fermionic excitations is increased continuously 
from zero to its full strength with the same rate for all excitation 
energies. 
This flow, also dubbed interaction flow, has been tested previously for the 2D 
Hubbard model without symmetry breaking and gives analogous results to 
those obtained with momentum-shell or temperature-flow schemes
\cite{2004PhRvB..70w5115H}. It can 
be regarded as an infinitely smooth cutoff, i.e. the opposite extreme to a 
sharp cutoff, for which all modes above a running energy scale 
$\Lambda$ are included with full weight and all modes below $\Lambda$ 
do not contribute.

Counterterms have been employed in the context of correlated
fermions before, 
notably in calculations for the symmetric state of interacting
electron systems.
The early work \cite{ft1} by Feldman and Trubowitz 
considers systems with spherical free Fermi
surfaces. Unphysical infrared divergencies of perturbation
theory Feynman graphs are removed by adjusting the chemical potential on the
propagator lines to the interacting Fermi surface,
countering this by an appropriate addition to the self-energy. 
The addition is determined by renormalization-group methods.
This procedure can be viewed as a shift from the chemical potential
into the interaction.
Feldman, Salmhofer, and Trubowitz \cite{fst1} 
employ it for more general Fermi surfaces
which necessitate a counterterm that is a function in momentum space.
They 
identify and find bounds for the most singular contributions
to this couterterm function under slightly more restrictive conditions
\cite{fst3}.
The authors prove regularity properties of the counterterm function
in \cite{fst2}, which they use in \cite{fst4}
to study the invertibility of the mapping induced by the
counterterm which transforms the interacting to the 
corresponding free Fermi surface. The analysis of
\cite{fst4} is restricted to space-inversion-symmetric dispersions.

Counterterms have also been used in the study of 
symmetry breaking. 
Feldman and Trubowitz \cite{ft2}
is devoted to flows with symmetry-breaking in
the Cooper channel for a spherically symmetric
electron-phonon system. Counterterms are again used to remove unphysical
divergencies as in \cite{ft1}. The BCS self-consistency equation is recovered
as an approximation from the flow. 
Neumayr and Metzner \cite{neumayr:035112} 
study the interplay of d-wave superconductivity
and Fermi surface deformations by a counterterm technique.
The effective interactions causing the symmetry breaking
are constructed order by order in perturbation theory.
The counterterm is determined by setting the self-energy
on the Fermi surface equal to zero, giving rise to a self-consistency
equation.

It is shown in
\cite{2004PThPh.112..943S,gersch:236} that such self-consistency
equations are also naturally contained in the fRG method applied here
and in 
\cite{2004PThPh.112..943S,2005EPJB...48..349G,gersch:236}.
This method is exact for the models considered, as is
the self-consistency treatment. 
However, it should be noted  that for general
models, self-consistent calculations just as fRG calculations are not
exact anymore and their results may differ. It may depend on the specific
case whether self-consistent calculations 
or the broader and more fundamental fRG
method give the more useful results.

The role of counterterms in the renormalization group literature
has focussed on the removal of divergencies. In this work,
the counterterm serves the purpose of selecting a certain
symmetry-broken configuration. It also opens a gap in the
spectrum of the system, which removes the divergencies
on the Fermi surface treated in the literature.

This paper is structured as follows. In section \ref{sec:model},
we introduce the model, presenting and explaining the mean-field
solution in section \ref{sec:mf}. 
We outline the renormalization-group scheme and
introduce our conterterm extension in section \ref{sec:RG}, where
we furthermore provide a numerical RG analysis of the model
from section \ref{sec:model}. We discuss our work and provide an
outlook in section \ref{sec:discussion}.


\section{Model}
\label{sec:model}

To illustrate the utility of the fermionic functional renormalization 
group (fRG) scheme introduced in section \ref{sec:RG},
we consider a tight-binding model on a $d$-dimensional 
hypercubic lattice with an interaction restricted to momentum transfers 
of $\vec{Q}=(\pi,\pi,\dots)$.
The Hamiltonian reads
\begin{eqnarray}
  H_{\rm red}=& \sum_{\vec{k}} \xi (\vec{k})
  \, c_{\vec{k}}^\dagger c^{\phantom\dagger}_{\vec{k}}
  -
  \frac{V_0}{N}\sum_{\vec{k}_1,\vec{k}_2}
  c^\dagger_{\vec{k}_1} c^{\phantom\dagger}_{\vec{k}_1+\vec{Q}}
  c^\dagger_{\vec{k}_2} c^{\phantom\dagger}_{\vec{k}_2-\vec{Q}}
  \nonumber\\
  &
  +
  \sum_{\vec{k}} 
  \left(
    \Delta_{\rm ext}+\Sigma_{\mathrm{i}}-\Delta_{\mathrm{c}}
  \right) \,
  c^\dagger_{\vec{k}} c^{\phantom\dagger}_{\vec{k}+\vec{Q}}
  .
  \label{eq:hamiltonian}
\end{eqnarray}
In this work, $\xi(\vec{k})= \epsilon (\vec{k}) - \mu$ includes a chemical
potential $\mu$, which can be adjusted to produce a system exhibiting
a first-order phase transition. We assume in the following that the 
dispersion fulfills the nesting relation 
$\epsilon (\vec{k}) =  -\epsilon (\vec{k}-\vec{Q})$. The kinetic energy $\xi (\vec{k})$, interaction, and external-field parts related to $ \Delta_{\rm ext}$ are also introduced and briefly discussed in Chapter 2 of \cite{2005EPJB...48..349G}.
For $V_0>0$ the interaction term can lead to charge-density-wave ordering 
with wavevector $\vec{Q}$. 
This will be discussed below employing a mean-field approach.
In fact, due to the restricted form of the interaction of the model, the mean-field treatment and the RG approach described below become exact in the thermodynamic limit, where the number of lattice sites $N \to \infty$ 
(for a diagrammatic argument, see \cite{gersch:236}). 
The same two-sublattice charge modulation is induced by the last term, which breaks the translational symmetry of the Hamiltonian by coupling an alternating charge field to the fermions.
In the fRG treatment, $\Sigma_{\mathrm{i}}$ will be used as
the initial condition for the off-diagonal self-energy. 
The counterterm $\Delta_{\mathrm{c}}$ will be included in the bare propagator 
to prevent this from affecting the physics of the system. 
We always set $\Sigma_{\mathrm{i}}=\Delta_{\mathrm{c}}$ which guarantees 
a cancellation at the end of the fRG flow.
$\Delta_{\rm ext}$ allows us to study the effect of an external field
and hysteresis phenomena.

\section{Mean-field treatment}
\label{sec:mf}
\begin{figure}[htbp]
  \centering
  \includegraphics[scale=.4]{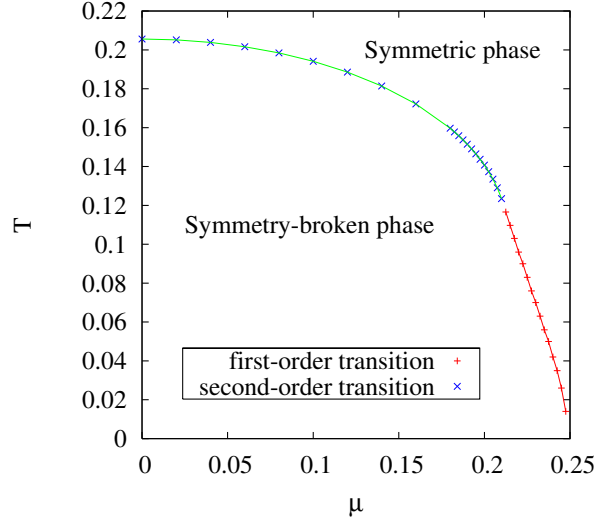}
  \caption{$\mu$-$T$ phase diagram of \eqref{eq:hamiltonian} for $V_0=2.0t$.}
  \label{fig:phasediagram}
\end{figure}

The low-temperature state of the model without explicit symmetry breaking, i.e. for
$\left( \Delta_{\rm ext}+\Sigma_{\mathrm{i}}-\Delta_{\mathrm{c}}
  \right)=0$, depends on the chemical potential $\mu$. 
For $\mu=0$ and 0.5 particles (per spin orientation) per lattice site, the Fermi surface is perfectly nested and mean-field theory finds a second-order transition toward the charge-density wave state at a critical temperature $T_c$. In the ordered state, 
$\afgap = \frac{V_0}{N} \sum_{\vec{k}} \langle  c^\dagger_{\vec{k}} c^{\phantom\dagger}_{\vec{k}+\vec{Q}} \rangle$ 
becomes nonzero.
Similarly, the RG without self-energy corrections would  find a runaway flow at a nonzero critical scale $\Lambda_c$. The runaway flow and the second-order mean-field transition are removed by a sufficiently large $\mu$ of the order of 
$T_c(\mu=0)$.  
For such $\mu$,
a local stability analysis of the symmetric state via an expansion around 
$\afgap = 0$ does not detect any instability. However, below 
a transition temperature $T_t$, a mean-field search reveals 
global minima of the thermodynamic potential as function of 
$\afgap$ which do not develop out of the local minimum 
at $\afgap =0$ when the temperature is lowered. 
As the temperature is lowered, the system jumps from 
$\afgap =0$ to the nonzero value, undergoing a first-order phase 
transition. The resulting phase diagram is shown in Fig.
\ref{fig:phasediagram}.
The thermodynamic potentials as functions of 
$\afgap$ for $\mu=0$ and for first-order cases with $\mu\not=0$ 
are shown in Fig. \ref{fig:MF_Om}.

\begin{figure}[htbp]
  \centering
  \includegraphics[scale=.4,clip=true, angle=-90]{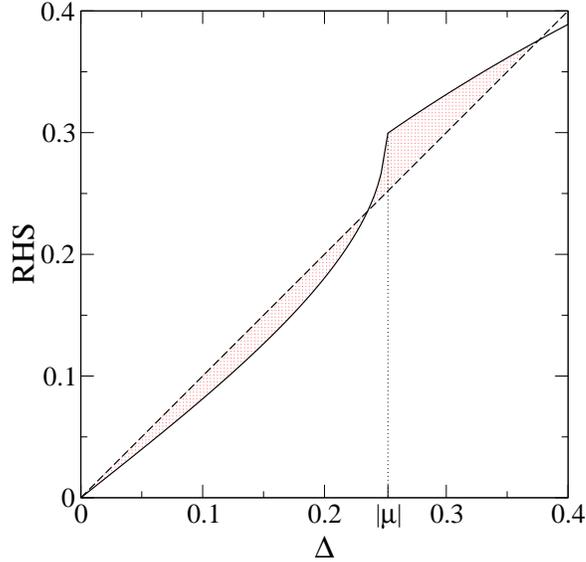}
  \caption{The right hand side of the gap equation (\ref{MF:self}) 
    has a kink at $\afgap=|\mu|$ leading to further solutions. 
    The solutions at $\afgap= 0 , 0.39t$ are minima while the
    solution at $\afgap=0.24t$ is a maximum of $\Omega$. 
    All data obtained for $V_0=2.0t$. }
  \label{fig:RHS}
\end{figure}

\begin{figure}[htbp]
  \centering
  \includegraphics[scale=.4,clip=true, angle=-90]{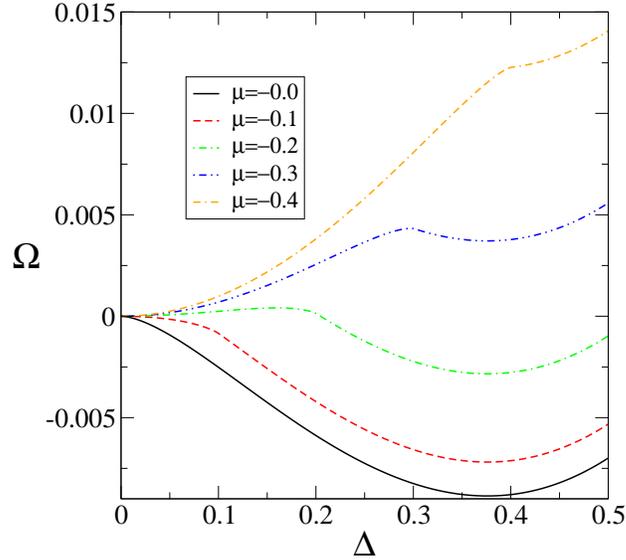}
  \caption{The thermodynamic potential for $t'=0$, $V_0=2.0t$. The position 
    of the minimum at non-zero gap values  does not change with $\mu$, 
    while the energy gain strongly depends
    on it. $\Omega(\afgap=0)$ is  subtracted from all curves.}
  \label{fig:MF_Om}
\end{figure}

To understand how such a simple model can have a first-order
transition, we  consider the self-consistency equation
\begin{equation}
\afgap = \frac{1}{N}\sum_{k} \frac{V_0 \afgap
}{E_k^+-E_k^{-}} \left(f(E_k^+)-f(E_k^-)\right), \label{MF:self}
\end{equation}
where $f(x)=(\exp(x/T)+1)^{-1}$ is the usual Fermi distribution and
the eigenvalues are 
\begin{equation}
E_k^\pm= \pm R -\mu,  \label{MF:EV}
\end{equation}
where $R=\sqrt{(\varepsilon_k)^2+\afgap^2}$. 
\eqref{MF:self} is obtained from
the differential of the grand canonical potential $\Omega$ 
by demanding $\partial_\afgap \Omega=0$.
To be specific, we base the following discussion on a 
two-dimensional square lattice with nearest neighbor
hopping $t$ (which is used as the unit of energy for the remainder of
the paper) at $T=0$. The self-consistency equation is
plotted in Fig.\ \ref{fig:RHS}. The RHS has a kink at $\afgap
=|\mu|$ because one of the branches of the eigenvalues
changes its sign on the whole umklapp surface for $\afgap=|\mu|$,
cutting out this part of the integration area of the
RHS.  It can be shown that at the kink
the slope diverges as $(1/\sqrt{|\mu|-\afgap})\log (|\mu|-\afgap)$ for $\afgap\nearrow|\mu|$.
At $T=0$, the RHS does not depend on $\mu$ for $\afgap>|\mu|$ since
$\mu$ enters the RHS only via the Fermi functions. Since $E_k^+$
($E_k^-$) is always positive (negative) for $\afgap>|\mu|$, 
the thermodynamically stable solution is always half filled in this case.
The dependence of the filling on $\afgap$ should not be neglected, as is sometimes done in the literature \cite{1985PhRvB..31.4403H}.
In fact, mean-field theory already predicts the breakdown
of the antiferromagnetic gap away from half filling
in the homogeneous solution, in agreement with the 
quantum Monte-Carlo calculations of \cite{1985PhRvB..31.4403H}.
For a detailed discussion, see \cite{juliusdiss}.
Since \eqref{MF:self} is obtained from $\partial_\afgap \Omega=0$, all
intersection points are extrema (or saddle-points) of $\Omega$. 
If all intersection points are extrema of $\Omega(\afgap)$,
the outermost one must be a minimum to guarantee the existence of
a stable solution.  By construction, the integral of the bisector minus the 
RHS of \eqref{MF:self}
is the energy gain through the opening of a finite gap. 
Thus, comparing the two shaded areas of Fig.\ \ref{fig:RHS}
determines which of the two minima given by the intersections 
has lower energy. 

Plots of the thermodynamic potential are shown in Fig.\
\ref{fig:MF_Om}. For small $|\mu|>0$, a local minimum develops at
$\afgap=0$, which becomes the global and finally the only minimum
if $\mu$ is increased further. 
Numerical study of \eqref{MF:self} and
$\Omega$ for various $T$ and $\mu$ leads to the phase diagram
of Fig. \ref{fig:phasediagram}


\section{Renormalization group}
\label{sec:RG}
In this section, we outline an extension of the one-particle irreducible (1PI)
fermionic functional renormalization group (fRG) scheme 
\cite{2001PThPh.105....1S}
with the goal of studying first-order phase transitions. 
We start from the
flow equations as suggested by Katanin \cite{2004PhRvB..70k5109K},
studied for the case of the breaking
of a continuous symmetry in \cite{2004PThPh.112..943S}
and for the case of the breaking of a discrete symmetry
in \cite{2005EPJB...48..349G}.
We use the interaction-flow method \cite{2004PhRvB..70w5115H}, 
which permits the fruitful employment of counterterms.
We show flows for first-order and second-order phase transitions,
discussing the effect of the counterterm on possible approximations.
We also study the effect of an external field on the flow, identifying a
criterion for unphysical starting points.

\subsection{Setup}
The basic setup of a 1PI fRG 
calculation with Katanin's modification is outlined 
for two different models in \cite{2004PThPh.112..943S, 2005EPJB...48..349G}.
To understand the problem solved in the following, we consider the
order-parameter dependence of the thermodynamic potential in the 
plot for $\mu=0$ in Fig. \ref{fig:MF_Om}.
In \cite{2004PThPh.112..943S,2005EPJB...48..349G},
a small external field is included as the initial condition for
the order parameter. This external field is
relevant in Wilson's fixed-point picture,
but does not appreciably change
the physics away from critical scales.
It biases the fRG calculation
toward the thermodynamic potential 
minimum at finite $\Delta$ (see Fig. \ref{fig:MF_Om} $\mu=0$), 
which becomes the endpoint of the flow.
This method is unable to deal with first-order phase transitions
where the thermodynamically stable state is separated from the
symmetric state by an energy barrier.
A typical thermodynamic potential as function of the order 
parameter is depicted in the plot for $\mu=-0.2t$ in Fig. \ref{fig:MF_Om}.
Because the external field must be small, the flow always veers 
toward the symmetric thermodynamic potential minimum. 
Other minima are inaccessible, even if their thermodynamic
potentials are smaller.
Furthermore, the method cannot be used to study hysteresis effects
appearing if systems exhibiting second-order phase transitions
(see the graphs for $\mu=0$ and $\mu=-0.1t$ in Fig. \ref{fig:MF_Om})
are placed in an external field.
The external field destroys the axis symmetry
of $\Omega(\Delta)$, but always biases the flow toward the global
minimum of the thermodynamic potential, making it impossible
to study the metastable configuration. 
Thus, the challenge encountered here is to set up an fRG scheme
with a parameter that allows the selection of any 
stable or metastable configuration as the endpoint of the flow.

The basic idea that will be exploited in the following is to
include in the calculation a counterterm $\Delta_c$ 
of arbitrary strength that cancels an equally strong
external field, but only at the end of the flow.
In the fRG calculation, the external field is taken
into account as the initial value $\Sigma_i$ of the order parameter 
$\Sigma$, while
the counterterm $\Delta_c$ is added to the naked propagator $\mathbb{Q}_0$.
Schematically, the matrix Green's function of the system reads
\begin{equation}
  \label{eq:countergreen}
  \mathbb{G}^{-1} = \frac{\mathbb{Q}_0+\Delta_c}{\chi}-\Sigma,
\end{equation}
where $\chi$ is the cutoff function and $\mathbb{Q}_0$ is the
inverse bare propagator without counterterm. For any $\chi \not\equiv 1$, counterterm and self-energy do not cancel each other. However, the special case $\chi\equiv 1$ is approached at the end of the RG flow. Thus, the vertex function of the symmetric model is obtained. 

We first consider the case of $\chi= \chi_\Lambda (\vec{k})$ describing the sharp
momentum-shell cutoff of conventional RG schemes (see \cite{PhysRevB.4.3184}). 
Then, only momenta for which $\chi_\Lambda (\vec{k})=1$ are taken into account in
the RG flow. For these momenta
$\Delta_c$ and $\Sigma$ cancel exactly at the start
of the flow, causing $\Sigma$ to remain stationary.
This is due to the right-hand side 
of the 1PI flow equation for the order parameter being 
proportional to the anomalous part of the Green's function 
which vanishes if $\Delta_c$ and $\Sigma$ cancel.
This argument generalizes to the case of multiple order parameters,
as can be seen by considering the adjugate matrix of the Green's function. 
Consequently, $\Delta_c$ and $\Sigma$ also cancel at all
stages of the flow. In the case of a second-order transition, the fRG will diverge at a nonzero scale since the spectrum remains unchanged. In the case of a first-order transition, the flow remains finite, signaling incorrectly the stability of the symmetric phase. We conclude that we have to employ
a softer cutoff function.

Next, we consider the softest possible cutoff function, provided by the
interaction flow scheme \cite{2004PhRvB..70w5115H}.
This scheme works by linearly turning on the interaction
with the flow parameter
(note the similarity to the treatment by Mahan \cite{mahan})
while turning on the counterterm
with the square root of the flow parameter. By rescaling the fields, 
this is found to be equivalent to employing a cutoff function
$\chi$ which is constant at all momenta, zero at the start of the flow,
and one at the end of the flow. With this, the counterterm
and the external field do not cancel in \eqref{eq:countergreen} until  
the end of the flow, and all configurations become accessible,
as will be illustrated in the following.

For a general cutoff function $\chi_\Lambda$ with flow parameter $\Lambda$ and
a general model with two-particle bare interaction, the
flow equations were obtained in \cite{2004PhRvB..70k5109K}:
\begin{equation}
\dot{\Sigma}^\Lambda = -\frac{1}{2}\Tr\left(\mathbb{V}^\Lambda\mathbb{G}^\Lambda
\frac{\partial}{\partial \Lambda}\left(\frac{\mathbb{Q}}{\chi_\Lambda}\right)
\mathbb{G}^\Lambda \right)
\label{eq:Sflow_general}
\end{equation}
\begin{equation}
  \label{eq:Vflow_general}
  \dot{{\mathbb V}}^\Lambda = -\frac{1}{2}\Tr\left(\mathbb{V}^\Lambda\frac{\partial}
    {\partial \Lambda}\left(\mathbb{G}^\Lambda\mathbb{G}^\Lambda\right)
    \mathbb{V}^\Lambda\right)+O(\mathbb{V}^3),
\end{equation}
where $\mathbb{G}^\Lambda=\left(\mathbb{Q}/\chi_\Lambda-\Sigma^\Lambda\right)^{-1}$
and the traces
run over all degrees of freedom of $\mathbb{G}^\Lambda$.
The contribution of $O(\mathbb{V}^3)$ to \eqref{eq:Vflow_general}
vanishes in the case of the Hamiltonian \eqref{eq:hamiltonian}
when taking the thermodynamic limit due to the restricted
momentum structure of the interaction.
For the interaction flow, the cutoff function is
\begin{equation}
  \label{eq:iacutoff}
  \chi_\Lambda:=\sqrt{\Lambda},
\end{equation}
where $\Lambda$ runs from zero to one.
Considering $\varepsilon=\xi+\mu$ as mapping the Brillouin
zone onto $\mathbb{R}$, denoting the effective gap 
$\Delta_{\mathrm f}:=-\Delta_c+\chi\Sigma$,
$R:=\sqrt{\varepsilon^2+\Delta_{\mathrm f}^2}$, and
$E^\pm:= -\mu\pm R$, 
we obtain the $\Lambda$-dependent inverse free propagator
\begin{equation}
  \label{eq:freelambda}
  \mathbb{Q}(\omega_n)=
  \frac{1}
       {\chi}
  \left(
  \begin{array}{cc}
    \mathrm{i} \omega_n-\varepsilon+\mu & \Delta_{c} \\ 
    \Delta_{c}            & \mathrm{i} \omega_n+\varepsilon+\mu
  \end{array}
  \right)
\end{equation}
and the $\Lambda$-dependent Matsubara Green's
function
\begin{eqnarray}
  \nonumber \fl
  \mathbb{G}(\omega_n)=
  \frac{\chi}
       {-\omega_n^2-(E^++E^-)\mathrm{i}\omega_n+E^+E^-}\times \\
  \left(
  \begin{array}{cc}
    \mathrm{i} \omega_n+\varepsilon+\mu & \Delta_{\mathrm{f}} \\ 
    \Delta_{\mathrm{f}}            & \mathrm{i} \omega_n-\varepsilon+\mu
  \end{array}
  \right).
  \label{eq:greenlambda}
\end{eqnarray}
We have omitted $\Lambda$-dependencies in the notation for brevity,
as we continue to do in the following.
\eqref{eq:freelambda} and \eqref{eq:greenlambda}
allow the calculation of the single-scale
propagator $\mathbb{S}=-\mathbb{G}\dot{\mathbb{Q}}\mathbb{G}$.
Tracing over all degrees of freedom, analytically evaluating
the Matsubara sum, and multiplying with the effective
interaction and diagrammatic factors, we obtain
the flow equation
\begin{eqnarray}
  \nonumber\fl
  \dot{\Sigma} = 
  -V\int\frac{\mathrm{d}^d k}
            {(2\pi)^d}     
  \frac{1}{4\sqrt{\Lambda}}
  \left\{
    \frac{f(E^-)-f(E^+)}{R}
    \left[
      \sqrt{\Lambda}\Sigma\frac{\xi^2}{R^2}+\Delta_{\mathrm{f}}
    \right] \right. \\
  \left.
    -\frac{f^\prime(E^-)+f^\prime(E^+)}{R^2}
      \sqrt{\Lambda}\Sigma\Delta_{\mathrm{f}}^2
  \right\}.\label{eq:sigmaflow}
\end{eqnarray}
Similarly treating the right-most diagram of Fig. 3 from 
\cite{2005EPJB...48..349G} or alternatively evaluating \eqref{eq:Vflow_general}
for the special case \eqref{eq:hamiltonian}
yields the interaction flow equation 
\begin{eqnarray}
  \nonumber\fl
  \dot{V} = 
  -V^2\int\frac{\mathrm{d}^d k}
            {(2\pi)^d}
  \frac{1}{2R^2}
  \left\{
    \frac{f(E^-)-f(E^+)}{R}
    \xi^2
    \left(
      3\Lambda\frac{\dot{R}}{R}
      -1
    \right)
  \right.
    +\\ \nonumber
    \left(
      f^\prime(E^-)+f^\prime(E^+)
    \right)
    \left(3\xi^2 \Lambda \frac{\dot{R}}{R} 
      +\Delta_{\mathrm{f}}^2
    \right)\\
    -
  \left.
    \vphantom{%
    \left(3\xi^2 \Lambda \frac{\dot{R}}{R} %
      +\Delta_{\mathrm{f}}^2%
    \right)%
    }
    \left(
      f^{\prime\prime}(E^-)+f^{\prime\prime}(E^+)
    \right)
    \Lambda\Delta_{\mathrm{f}}^2\dot{R}
  \right\}.
  \label{eq:Vflow}
\end{eqnarray}

We furthermore need to calculate the thermodynamic potential using
the functional renormalization group. From 
equation (48) of \cite{2001PThPh.105....1S}, 
using $\Omega_{\mathrm{i.a.}}=T\gamma_0$ and equations (39) and (45) of 
\cite{2001PThPh.105....1S},  we determine the flow equation 
\begin{eqnarray}
  \nonumber
  \dot{\Omega} &=\frac{T}{2}\Tr
  \left(
    \left(
      \mathbb{G}-{\mathbb{Q}}^{-1}
    \right)
    \dot{\mathbb{Q}}
  \right)\\
  \nonumber
  &=-\frac{T}{2}\Tr
  \left(
    \frac{\dot{\chi}}{\chi}
    \left(
      1+{\mathbb{Q}}^{-1}\Sigma+
      \left({\mathbb{Q}}^{-1}\Sigma\right)^2+\dots
    \right)
    -\frac{\dot{\chi}}{\chi}
  \right) \\
  &=-\frac{T}{2}\Tr
  \left(
    \Sigma\frac{\dot{\chi}}{\chi}\mathbb{G}
  \right).  \label{eq:omegaflowcalc}
\end{eqnarray}
If the flow is started at the thermodynamic potential
of the non-interacting system, the full thermodynamic
potential is recovered. Since we will always do so,
we have written $\Omega$ instead of $\Omega_{\mathrm{i.a.}}$
in \eqref{eq:omegaflowcalc}.
The right-hand side of \eqref{eq:omegaflowcalc} can again be drawn
as a 1PI diagram.
Evaluating the Matsubara sums,
\eqref{eq:omegaflowcalc} reads
\begin{equation}
  \label{eq:omegaflow}
  \dot{\Omega}=\int\frac{\mathrm{d}^d k}
            {(2\pi)^d}
  \Sigma\Delta_{\mathrm{f}}
  \frac{f(E^-)-f(E^+)}{4R\sqrt{\Lambda}}.
\end{equation}
\eqref{eq:sigmaflow}, \eqref{eq:Vflow}, and \eqref{eq:omegaflow}
constitute a closed system of integro-differential equations
which can be numerically solved given appropriate 
initial conditions.

\subsection{Flows for first-order phase transitions}
\begin{figure}[htb]
  \includegraphics[scale=.4]{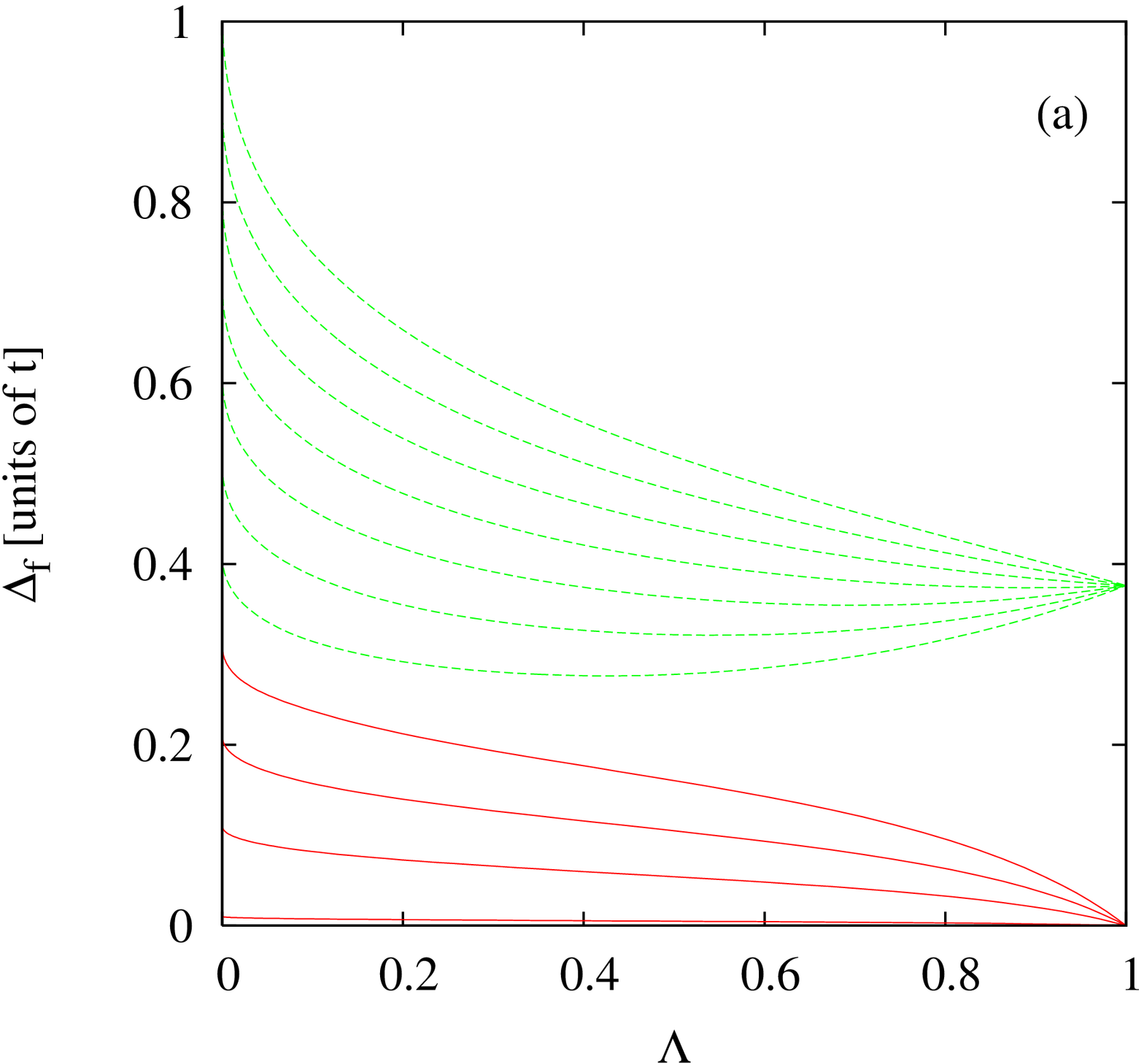}
  \includegraphics[scale=.4]{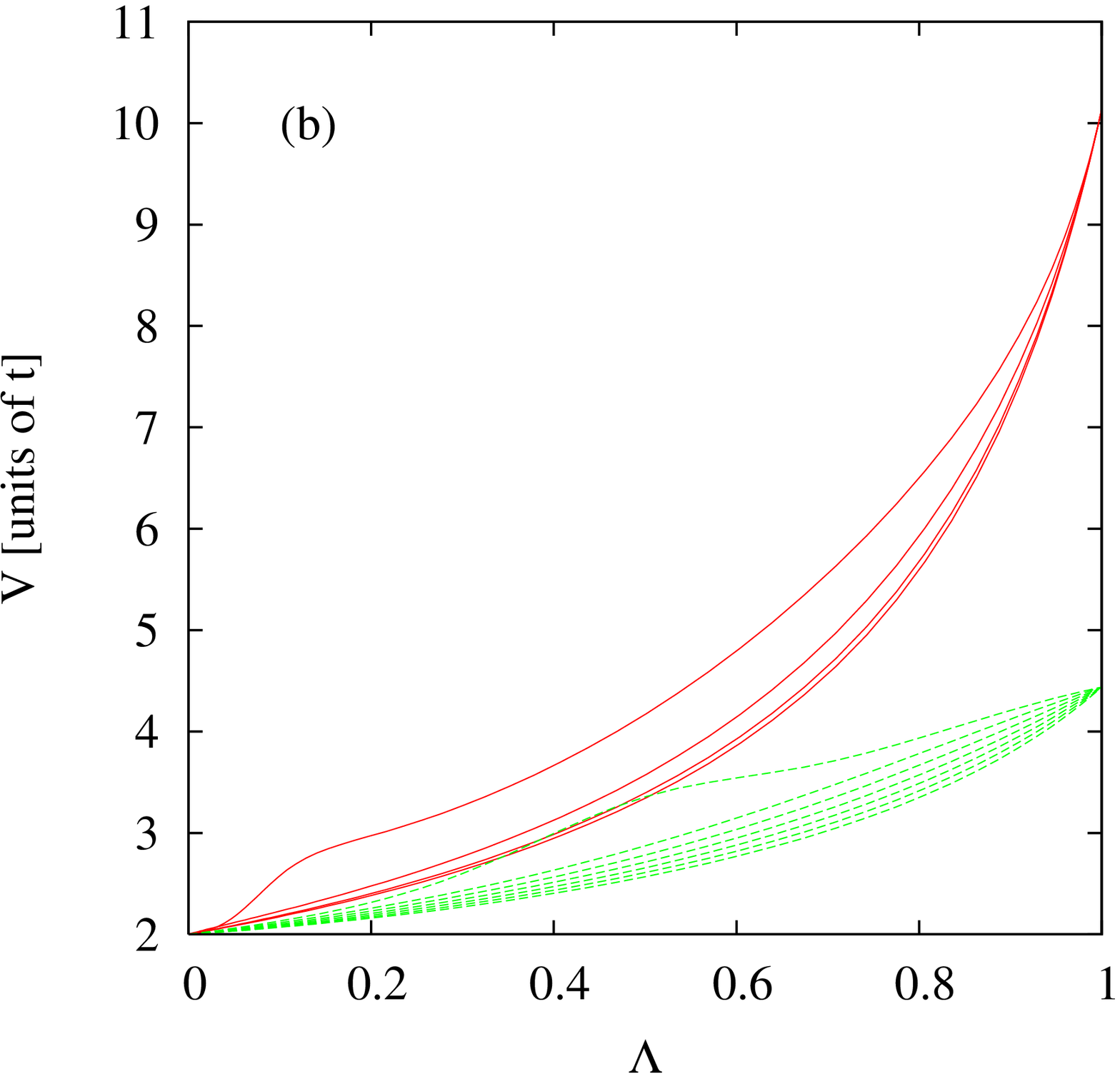}
  \includegraphics[scale=.4]{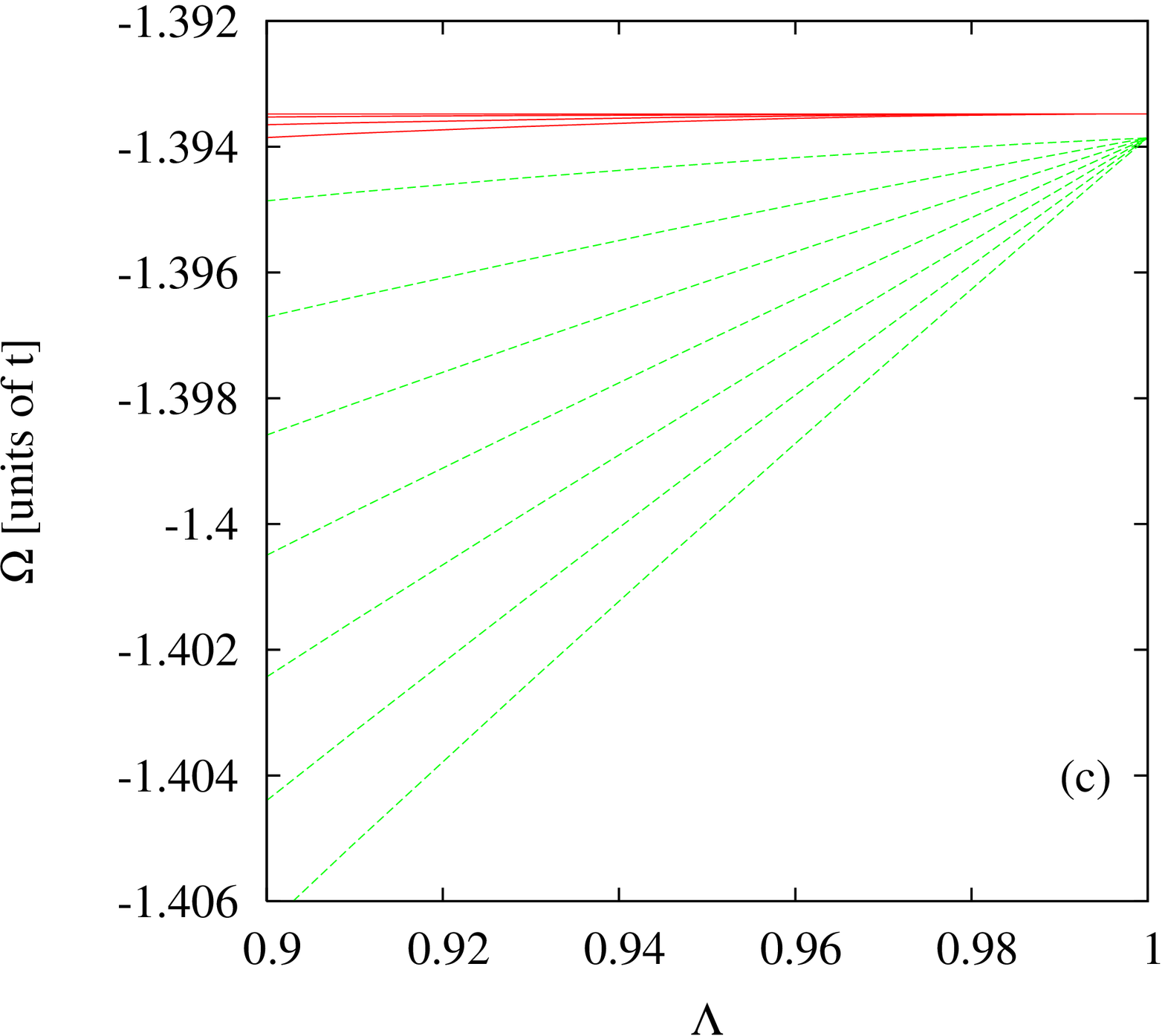}
  \caption{Flows for $V_0=2t$ at $\mu=0.245t$, $T=0.01t<T_t$, and 
    $\Delta_c$ increasing from $0.01t$ to $1.01t$
    in increments of $0.1t$. Broken lines denote flows converging to
    the stable symmetry-broken configuration; solid lines denote
    flows converging to the metastable symmetric configuration.
    The flow starts at $\Lambda =0$ and finishes at $\Lambda=1$.
    (a) Effective gap $\Delta_{\mathrm{f}}$. $\Delta_c$ for
    each graph can be read off at the y-axis.
    (b) Effective interaction $V$. Flows converging at $10.14t$
    pass through greater values with increasing 
    counter-term. Flows converging at $4.43t$ behave inversely.
    (c) Thermodynamic potential $\Omega$. Flows
    pass through smaller values with increasing counterterm.}
  \label{fig:lowTflow1st}
\end{figure}

We first consider a system exhibiting a first-order phase transition at a transition temperature $T_t$.
Studying flows below $T_t$ (see Fig. \ref{fig:lowTflow1st}), we notice two strong attractors. By its lower
thermodynamic potential at the end of the flow, one of them can be identified with the stable, symmetry-broken configuration. 
The values for the order parameter, effective
interaction, and thermodynamic potential difference reproduce the
exact mean-field results. Note that the final values do not depend on
the magnitude of the counterterm. The dependence of the results
on the external field, known from \cite{2005EPJB...48..349G,2005EPJB...48..319D,2004PThPh.112..943S}, 
is eliminated.

Considering Fig. \ref{fig:lowTflow1st}(a), we see that there is a separatrix 
between the effective gap flows to the metastable
configuration and the flows to the stable configuration.
For more complicated problems than \eqref{eq:hamiltonian},
the flow probably has to be stopped at $\Lambda<1$ because
of the rise of large effective interactions which are not correctly
taken into account by our method. Therefore, flows which exhibit
weak changes of the effective interaction away from $\Lambda=1$
and weak changes of the order parameter close to $\Lambda=1$
are most promising for determining order parameters
in more complicated models.
We see that the flows closest to the separatrix are not optimal
flows since $\Delta(\Lambda)$ still shows a large slope at $\Lambda=1$.
This final slope is smallest for a counterterm of roughly
twice the magnitude of the order parameter. Such a flow would
yield an excellent approximation of the order parameter
even if stopped at $\Lambda=0.75$.

\begin{figure}[htb]
  \includegraphics[scale=.4]{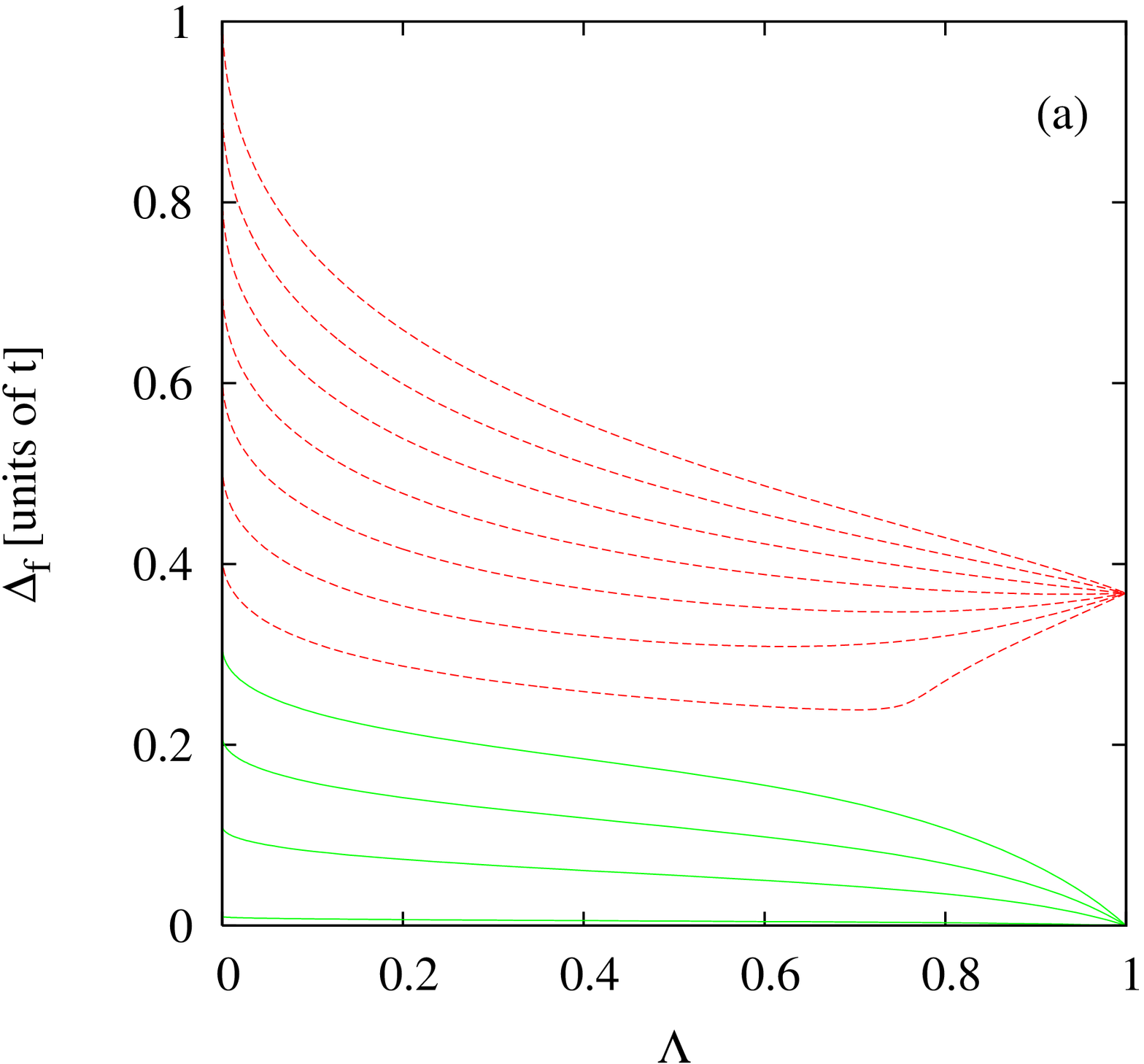}
  \includegraphics[scale=.4]{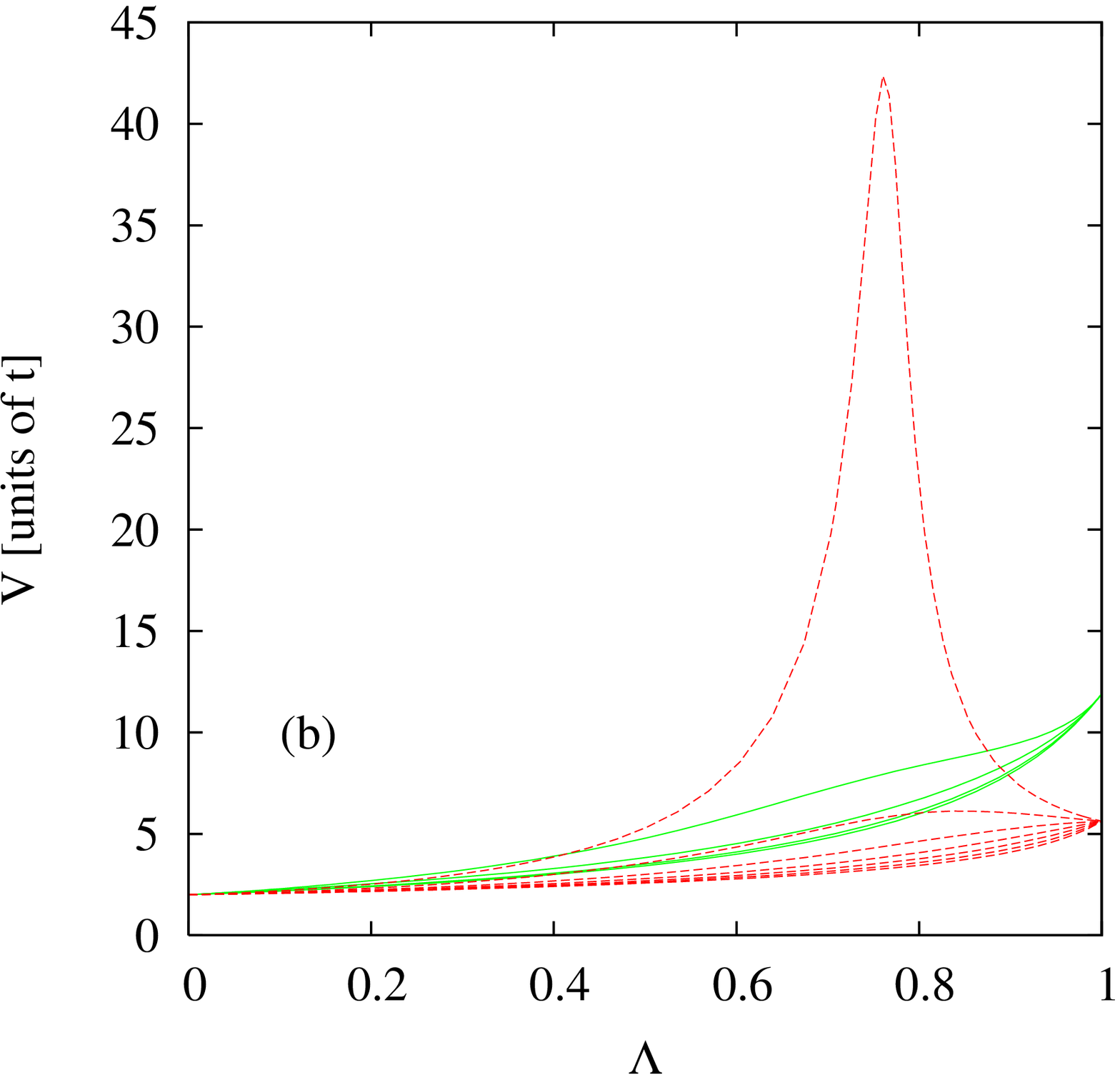}
  \includegraphics[scale=.4]{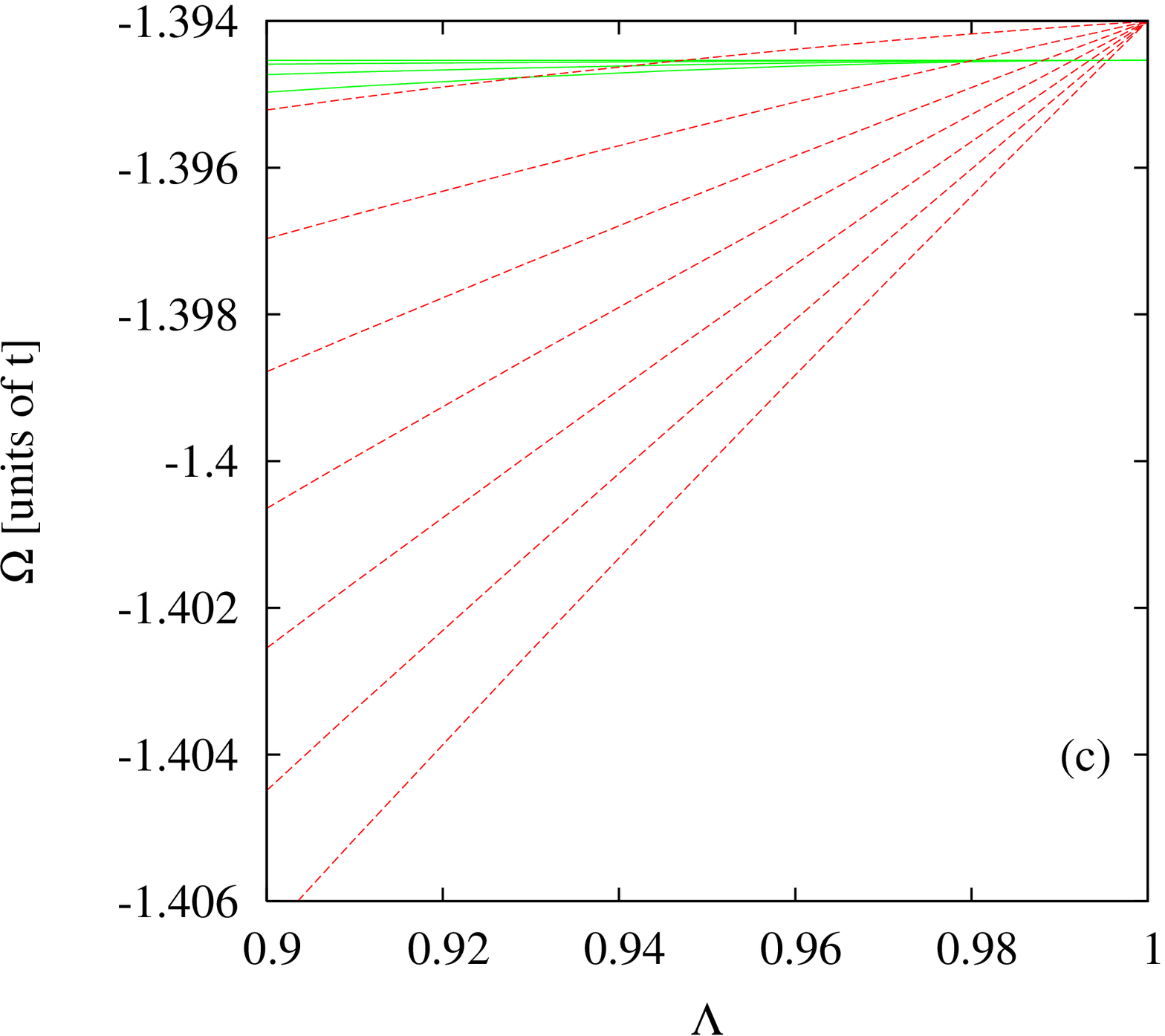}
  \caption{Flows for $V_0=2t$ at $\mu=0.245t$, $T=0.04t>T_t$, and 
    $\Delta_c$ increasing from $0.01t$ to $1.01t$
    in increments of $0.1t$. Broken lines denote flows converging to
    the metastable symmetry-broken configuration; solid lines denote
    flows converging to the stable symmetric configuration. The flow starts at $\Lambda =0$ and finishes at $\Lambda=1$.
    (a) Effective gap $\Delta_{\mathrm{f}}$. $\Delta_c$ for
    each graph can be read off at the y-axis.
    (b) Effective interaction $V$. Flows converging at $11.90t$
    pass through greater values with increasing 
    counter-term. Flows converging at $5.62t$ behave inversely.
    (c) Thermodynamic potential $\Omega$. Symmetric flows pass
    through smaller values with increasing counterterm, and so
    do symmetry-broken flows.}
  \label{fig:highTflow1st}
\end{figure}

Considering the flows of the effective interaction
in diagram \ref{fig:lowTflow1st}(b), we do not find a separatrix
as in (a). Instead, the flows corresponding to the ones
close to the separatrix in (a) develop a shoulder which becomes
a maximum for certain values of the counterterm (for an
example of such a maximum, see diagram \ref{fig:highTflow1st}(b)).
The flows exhibiting the strongest effective interactions
correspond to the flows giving the worst approximations for
the order parameter if terminated prematurely.
This entails for applications to models where this method is not exact
that carefully chosing the counterterm can significantly simplify
the calculation and improve the accuracy of the results.

The flows of the thermodynamic potential shown in \ref{fig:lowTflow1st}(c)
again exhibit separatrix behaviour. This is in
contrast to the flows above $T_t$. For the thermodynamic potential, the
flows closest to the separatrix yield the best approximations
if terminated prematurely. However, it is apparent from 
\ref{fig:lowTflow1st}(c) that the flow must
be continued until the external field is completely compensated
by the counterterm to obtain reliable results for the thermodynamic 
potential.

Flows above $T_t$ are illustrated in Fig. \ref{fig:highTflow1st}.
Again, we clearly discern two attractors. In this case, however,
the thermodynamic potential flows of the symmetry-broken-phase
attractor cross the thermodynamic potential flows
of the symmetric-phase attractor. The symmetric phase is therefore
thermodynamically more stable. Apart from this, the flows
behave similarly as those below $T_t$. For inconveniently chosen
counterterms, the maxima in the effective-interaction flows
are clearly visible in \ref{fig:highTflow1st}(b). Nevertheless,
the flows reproduce the exact mean-field results.

\subsection{Flows for second-order phase transitions}
\begin{figure}[htb]
  \includegraphics[scale=.4]{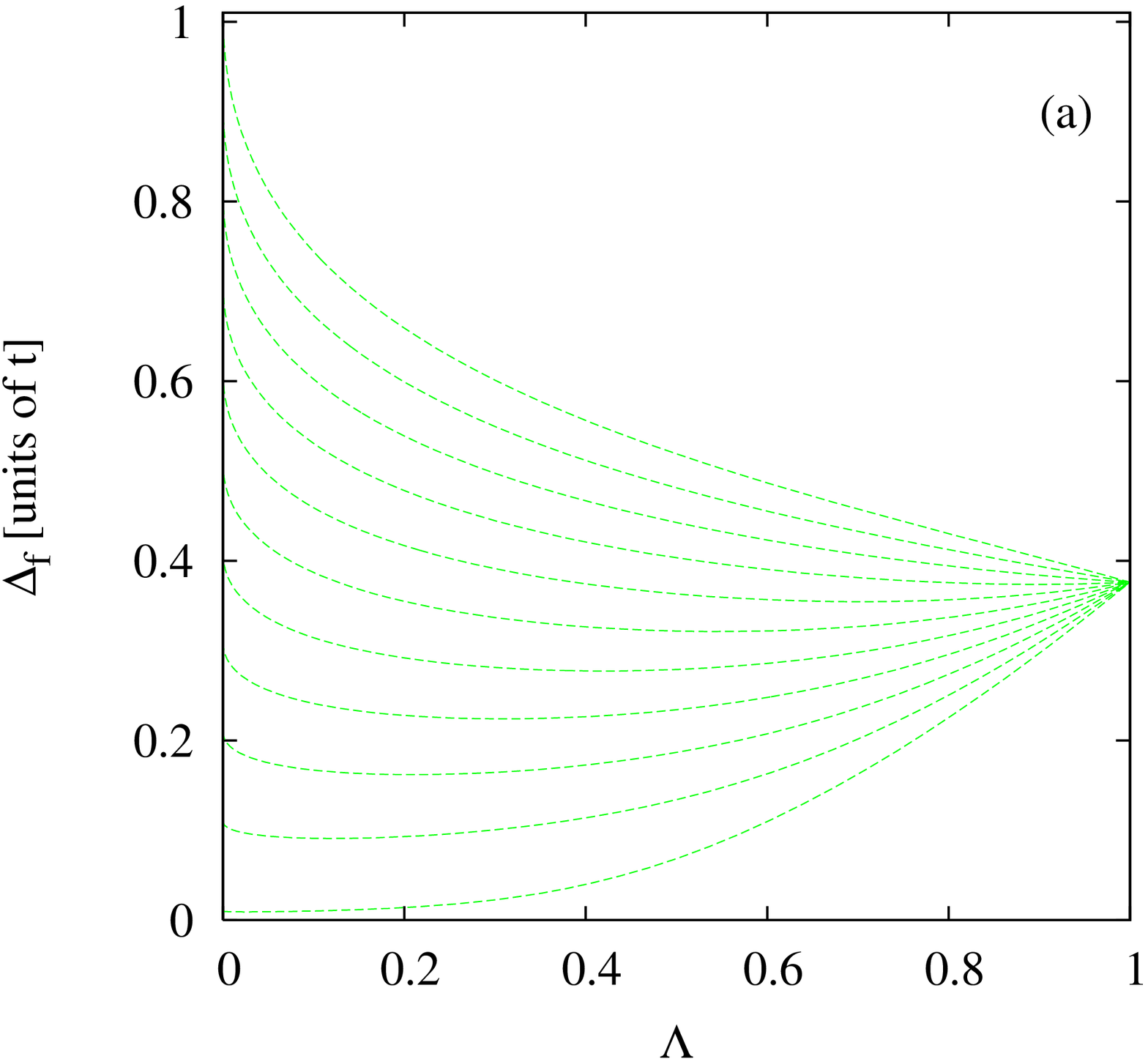}
  \includegraphics[scale=.4]{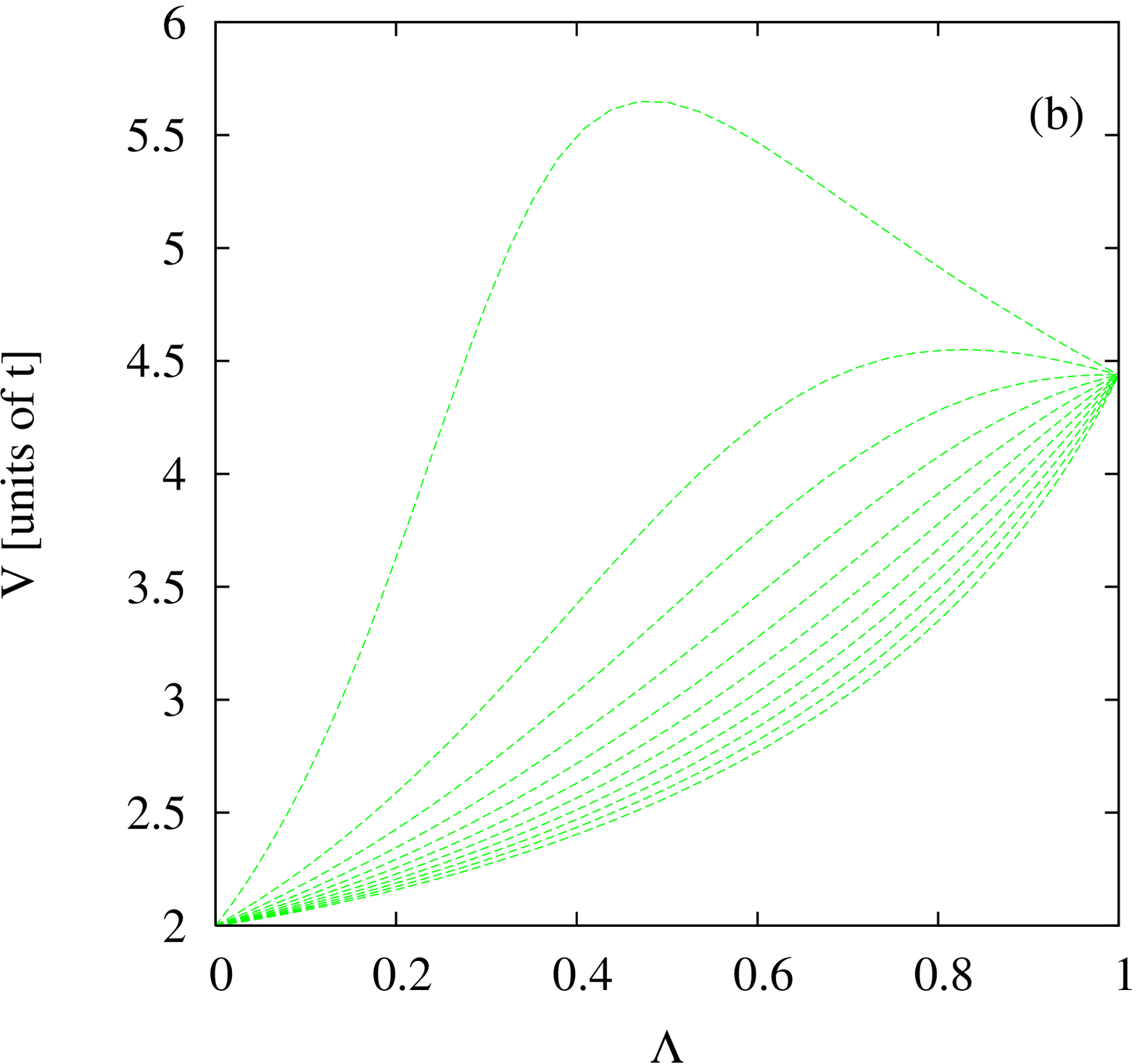}
%
  \caption{Flows for $V_0=2t$ at $\mu=0$, $T<<T_c$, and $\Delta_c$ increasing from $0.01t$ to $1.01t$ in increments of $0.1t$. Broken lines denote flows converging to the stable symmetry-broken configuration. The flow starts at $\Lambda =0$ and finishes at $\Lambda=1$. (a) Effective gap $\Delta_{\mathrm{f}}$. $\Delta_c$ for
    each graph can be read off at the y-axis.
    (b) Effective interaction $V$. Flows pass through smaller 
    values with increasing counterterm. 
  }
  \label{fig:lowTflow2nd}
\end{figure}

The flows for second-order phase transitions behave similarly to the
flows for first-order phase transitions, but only a single 
attractor appears, as can be seen in Fig. \ref{fig:lowTflow2nd}(a). 
Fig. \ref{fig:lowTflow2nd}(b) shows that it is possible 
to suppress the effective interaction during the flow
by chosing a large counterterm. Approximations for
the effective gap can be obtained by stopping the flow before
$\Lambda$ reaches $1$. The quality of such an approximation
depends on the counterterm chosen. Again, the best approximation
can be obtained by chosing a counterterm of 
twice the non-approximated value of the gap.

\subsection{External field}
If the external field $\Delta_{\mathrm{ext}}$ in \eqref{eq:hamiltonian}
is zero, the half-filled system below $T_c$ exhibits two degenerate
stable configurations (local minima of the thermodynamic
potential) distinguished by the sign of the
order parameter.
If $\Delta_{\mathrm{ext}}$ is non-zero, this degeneracy is lifted.
The fRG scheme outlined above allows us to select the endpoint of the flow independently
of the external field by appropriately setting the counterterm, in contrast to the
scheme from \cite{2004PThPh.112..943S,2005EPJB...48..349G}.
\begin{figure}[htbp]
  \centering
  \includegraphics[scale=.4]{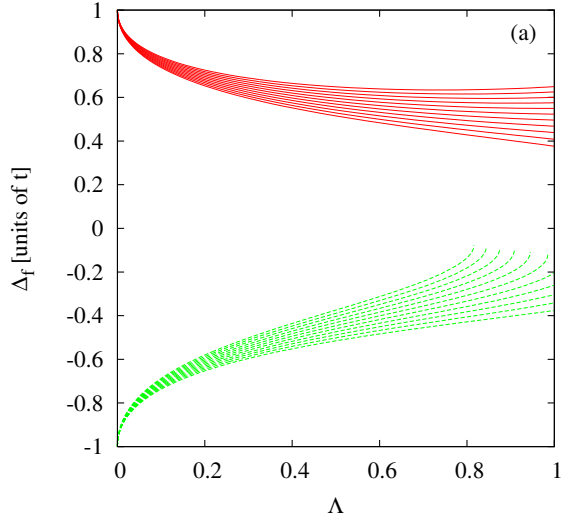}
  \caption{Flows of $\Delta_{\mathrm{f}}$ at $\mu=0$ for $V_0=2t$. Full
    (broken) lines have $\Delta_{\mathrm{c}}=(-)t$. The external
    field increases from $0$ to $0.15t$ with increasing final
    values of the flow.
    The flow starts at $\Lambda =0$ and finishes at $\Lambda=1$.}
  \label{fig:hysteresis_flows}
\end{figure}
The flows toward the metastable configuration
are shown in Fig. \ref{fig:hysteresis_flows} (broken lines).
As the external field is increased toward a critical value,
the order parameter for the metastable configuration
vanishes and the corresponding local minimum of the
thermodynamic potential disappears.
Beyond this critical value, the flows for negative
counterterms become divergent as the effective gap
reaches values close to zero, exposing low-energy modes.
Such a divergence can be used as an indicator for
an adversely chosen starting point.
The flows for positive counterterms 
(full lines in Fig. \ref{fig:hysteresis_flows}) always attain the
stable solution. Calculating the flows as for Fig. \ref{fig:hysteresis_flows},
but for negative values of the external
field $\Delta_{\rm ext}$, we obtain the hysteresis
curve of Fig. \ref{fig:hysteresis}.
\begin{figure}[htbp]
  \centering
  \includegraphics[scale=.4]{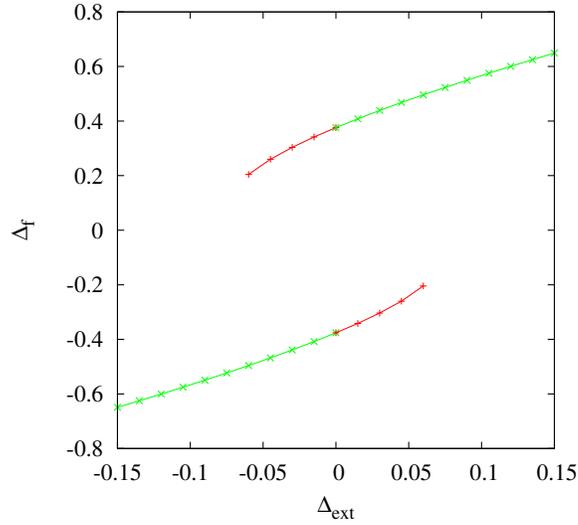}
  \caption{Hysteresis effect in an external field. All solutions 
    represent symmetry-broken configurations. Green lines ($\times$-symbols) 
    represent stable, red lines ($+$-symbols) metastable configurations.}
  \label{fig:hysteresis}
\end{figure}


\section{Discussion and outlook}
\label{sec:discussion}

We have described a counter-term scheme within the framework of the 
functional renormalization group concept for interacting fermions. The 
scheme is able to detect ordered ground states which, in terms of the 
effective bosonic potential for the ordering field, occur as first-order 
transitions. Second-order transitions are captured as well.  Therefore, 
the new scheme remedies one shortcoming of conventional 
functional RG 
techniques (e.g.,  \cite{2000PhRvB..6113609Z,2000PhRvL..85.5162H,2001PhRvB..63c5109H}) for interacting fermions which are blind 
with respect to symmetry-broken states which are separated by an energy 
barrier from the symmetric state. A second shortcoming that
is addressed is the need for a finite external field 
\cite{2004PThPh.112..943S,2005EPJB...48..349G}, which does not
arise in our approach. As a third improvement,
at least for discrete-symmetry breaking away from the critical
temperature, our scheme can circumvent the development of 
large interactions in the flow.

The validity of the new scheme has been demonstrated in a mean-field 
model, where our truncated version of the fRG equations and mean-field 
theory are known to be exact.  More generally, the scheme can be used to 
target individual minima of the effective bosonic potential, for example 
to study metastable states in the presence of an external field. As
an example, we have shown the calculation
of a hysteresis curve.

The fRG scheme can be applied to more realistic models such as the Hubbard 
model as well. Here a new complication arises, as the truncated form of 
the RG equations used is no longer exact. Hence large values of the 
coupling constants could spoil the validity. This is a general problem 
for perturbative functional renormalization group schemes which can for example be circumvented by partial bosonization at the expense of introducing a 
dynamical bosonic field 
\cite{2004PhRvB..70l5111B,2006JPhA...39.8205S,krahlwetterich,2005PhRvB..72c5107S}.
The counter-term technique presented here allows to avoid this problem 
in certain cases. As demonstrated and described above, in the case of 
the breaking of a {\em discrete symmetry}, no large couplings occur 
unless one is close to a curvature change of the effective bosonic 
potential. Hence for these cases, an inclusion of non-mean-field-type 
interactions should be possible without much trouble. For broken {\em 
continuous symmetries}, Goldstone modes may again lead to large couplings. 
It will be interesting to see whether the Goldstone physics 
only enters the flow in a critical way toward the end of the flow, 
when the off-diagonal selfenergy has already converged to the final 
value. The results above indicate a rather quick convergence for most 
values of the counter-term. These issues will be the subject of future work.

We emphasize however that even in the case that the flow cannot be followed 
over the full range, the present scheme should be useful for the 
detection of symmetry-broken states which are separated by an energy barrier 
from the symmetric state. For example, if the conventional analysis 
signals a locally stable symmetric state, the counter-term technique can 
be used to scan non-zero values of the suspected order parameter. If 
there is a lower thermodynamic potential minimum further out, the self-energy 
shows a flow toward non-zero values. Further, there should be a 
local curvature change of the effective bosonic potential somewhere 
between 0 and the value related to the global minimum, which is detected 
as a runaway flow of the interactions in the corresponding channel. 
Hence, the method is useful at least for qualitative scans of the 
bosonic potential landscape.
More generally, variations of the proposed scheme can be used for the perturbative analysis around states which differ from the free ground state of the model. This as well should lead to numerous new applications.  

\ack We thank T. Enss, W. Metzner, A. Katanin, M. Salmhofer, and P. Strack for pleasant discussions.
\\[1cm]

\bibliography{main}

\begin{thebibliography}{10}

\bibitem{PhysRevB.4.3184}
K.~G. Wilson, \prb {\bf 4},  3184  (1971).

\bibitem{1994RvMP...66..129S}
R. {Shankar}, \rmp {\bf 66},  129  (1994).

\bibitem{SalmhoferBook}
M. {Salmhofer}, Springer Texts and Monographs in Physics, Springer, Heidelberg
  (1998).

\bibitem{2001PThPh.105....1S}
M. {Salmhofer} and C. {Honerkamp}, \ptp {\bf 105},  1  (2001).

\bibitem{1993PhLB..301...90W}
C. {Wetterich}, \plb {\bf 301},  90  (1993).

\bibitem{2004PhRvB..70l5111B}
T. {Baier}, E. {Bick}, and C. {Wetterich}, \prb {\bf 70},  125111  (2004).

\bibitem{2006JPhA...39.8205S}
F. {Sch{\"u}tz} and P. {Kopietz}, Journal of Physics A Mathematical General
  {\bf 39},  8205  (2006).

\bibitem{krahlwetterich}
H.~C. Krahl and C. Wetterich, cond-mat/0608667  (2006).

\bibitem{2004PhRvB..70k5109K}
A.~A. {Katanin}, \prb {\bf 70},  115109  (2004).

\bibitem{1957PhRv..108.1175B}
J. {Bardeen}, L.~N. {Cooper}, and J.~R. {Schrieffer}, \pr {\bf 108},  1175
  (1957).

\bibitem{2004PThPh.112..943S}
M. {Salmhofer}, C. {Honerkamp}, W. {Metzner}, and O. {Lauscher}, \ptp {\bf
  112},  943  (2004).

\bibitem{2005EPJB...48..349G}
R. {Gersch}, C. {Honerkamp}, D. {Rohe}, and W. {Metzner}, European Physical
  Journal B {\bf 48},  349  (2005).

\bibitem{reissrohemetzner}
W. {Metzner}, J. {Reiss}, and D. {Rohe}, \pssb {\bf 243},  46  (2006).

\bibitem{juliusdiss}
J. Reiss, Ph.D. thesis, Universit\"{a}t Stuttgart, 2006.

\bibitem{2005PhRvB..72c5114Y}
H. {Yamase}, V. {Oganesyan}, and W. {Metzner}, \prb {\bf 72},  035114  (2005).

\bibitem{2004PhRvB..70w5115H}
C. {Honerkamp}, D. {Rohe}, S. {Andergassen}, and T. {Enss}, \prb {\bf 70},
  235115  (2004).

\bibitem{ft1}
J. {Feldman} and E. {Trubowitz}, \hpa {\bf 63},  156  (1990).

\bibitem{fst1}
J. {Feldman}, M. {Salmhofer}, and E. {Trubowitz}, \jsp {\bf 84},  1209  (1996).

\bibitem{fst3}
J. {Feldman}, M. {Salmhofer}, and E. {Trubowitz}, \cpam {\bf 52},  273  (1999).

\bibitem{fst2}
J. {Feldman}, M. {Salmhofer}, and E. {Trubowitz}, \cpam {\bf 51},  1133
  (1998).

\bibitem{fst4}
J. {Feldman}, M. {Salmhofer}, and E. {Trubowitz}, \cpam {\bf 53},  1350
  (2000).

\bibitem{ft2}
J. {Feldman} and E. {Trubowitz}, \hpa {\bf 64},  213  (1991).

\bibitem{neumayr:035112}
A. {Neumayr} and W. {Metzner}, Physical Review B (Condensed Matter and
  Materials Physics) {\bf 67},  035112  (2003).

\bibitem{gersch:236}
R. Gersch, C. Honerkamp, D. Rohe, and W. Metzner,  in {\em Fermionic
  Renormalization Group Flow at All Scales: Breaking a Discrete Symmetry},
  edited by F. Mancini and A. Avella (AIP, Melville, New York, 2006), No.~1,
  pp.\ 236--244.

\bibitem{1985PhRvB..31.4403H}
J.~E. {Hirsch}, \prb {\bf 31},  4403  (1985).

\bibitem{mahan}
G.~D. Mahan,  in {\em Many-Particle Physics}, {\em Physics of Solids and
  Liquids}, third edition ed. (Kluwer Academic/Plenum Publishers, New York,
  2000), Chap.~3, pp.\ 148--152.

\bibitem{2005EPJB...48..319D}
N. {Dupuis}, European Physical Journal B {\bf 48},  319  (2005).

\bibitem{2000PhRvB..6113609Z}
D. {Zanchi} and H.~J. {Schulz}, \prb {\bf 61},  13609  (2000).

\bibitem{2000PhRvL..85.5162H}
C.~J. {Halboth} and W. {Metzner}, Physical Review Letters {\bf 85},  5162
  (2000).

\bibitem{2001PhRvB..63c5109H}
C. {Honerkamp}, M. {Salmhofer}, N. {Furukawa}, and T.~M. {Rice}, \prb {\bf 63},
   035109  (2001).

\bibitem{2005PhRvB..72c5107S}
F. {Sch{\"u}tz}, L. {Bartosch}, and P. {Kopietz}, \prb {\bf 72},  035107
  (2005).

\end{thebibliography}
\end{document}